\documentclass{PoS}
\pdfoutput=1

\usepackage{amsmath}
\usepackage{amssymb}
\usepackage{dsfont}
\usepackage{tabularx}
\usepackage{multirow}
\usepackage{wrapfig,lipsum,booktabs}
\usepackage{tikz}
\usetikzlibrary{arrows,shapes}
\usetikzlibrary{decorations.markings}
\usepackage{cite}
\title{Mesons upon low-lying Dirac mode exclusion}

\ShortTitle{Mesons upon low-lying Dirac mode exclusion}

\author{\speaker{M. Denissenya}\\
        Institut f\"ur Physik, FB Theoretische Physik, Universit\"at Graz\\
        E-mail: \email{mikhail.denissenya@uni-graz.at}}

\author{L.~Ya.~Glozman\\
        Institut f\"ur Physik, FB Theoretische Physik, Universit\"at Graz\\
        E-mail: \email{leonid.glozman@uni-graz.at}}
        
\author{C.~B.~Lang\\
        Institut f\"ur Physik, FB Theoretische Physik, Universit\"at Graz\\
        E-mail: \email{christian.lang@uni-graz.at}} 

\abstract{We study  the isoscalar and isovector $J=0,1$ mesons  with the overlap operator within 
two flavour lattice QCD. After subtraction of the lowest-lying Dirac
eigenmodes from the valence quark propagator all disconnected contributions vanish and all possible
point-to-point $J=0$ correlators become identical, signaling 
 a simultaneous restoration of both $SU(2)_L \times SU(2)_R$ and $U(1)_A$ 
 symmetries. The ground states of the $\pi,\sigma,a_0,\eta$ mesons do not survive
 this truncation. All possible $J=1$ states have a very clean
 exponential decay and become degenerate, demonstrating a $SU(4)$ symmetry of a
 dynamical QCD-like string.}

\FullConference{The 32nd International Symposium on Lattice Field Theory,\\
		23-28 June, 2014\\
		Columbia University New York, NY}

\begin{document}
\section{ Introduction} 

in our previous study \cite{DGL} we discovered a  degeneracy of all isovector mesons of  spin $J=1$, $\rho, \rho',
a_1, b_1$, upon truncation of the quasi-zero modes from the valence quark
propagators with the manifestly chirally-invariant
overlap Dirac operator ( for a previous study with the chirally improved
Dirac operator see Refs. \cite{LS,GLS}). The density
of the quasi-zero modes is directly related to the quark condensate
of the vacuum \cite{BC}. 
Via such a truncation we
artificially restore ( "unbreak") the chiral symmetry.
All $J=1$ states survive the unbreaking, because a very clean exponential decay
of the correlators is seen. Surprisingly, not
only a degeneracy within the chiral partners $\rho$ and $a_1$ has been
observed, but actually a degeneracy of all four states. This degeneracy
signals not only a simultaneous restoration of both $SU(2)_L \times SU(2)_R$ 
and $U(1)_A$  symmetries, but  of some higher symmetry that
includes $SU(2)_L \times SU(2)_R \times U(1)_A$ as a subgroup. This symmetry
would require a degeneracy of all possible chiral multiplets with $J=1$
that contain both isoscalar and isovector mesons. Here
we report  our results on both the isoscalar and isovector mesons \cite{DGL2}.

Our second aim is to investigate the $J=0$ correlators in both the isovector
and isoscalar channels and clarify the fate of the $\pi,\sigma,a_0,\eta$
mesons upon unbreaking.

\section{ Lattice setup}

Our ensemble includes $100$ gauge configurations generated by JLQCD
with $n_f=2$ dynamical overlap fermions
at fixed topology $Q_{top}=0$ \cite{Aoki:2009qn}. The lattice size is $16^3\times 32$, the lattice
spacing $a\sim 0.12$ and the pion mass in this ensemble is $M_{\pi} =289(2)$ MeV. 

\subsection{Quark propagators}

The quark propagators, obtained from JLQCD, consist of two parts. The
contribution of the first 100 Dirac eigenmodes was computed exactly, and
the effect of all higher-lying modes was taken into account via a stochastic
estimate \cite{Aoki:2012pma}. Our unbreaking procedure means consequently a
removal of the first $k$ modes from the quark propagators:
\begin{equation}
 S_{k}(x,y)=\sum^{100}_{n=k+1}\frac{1}{\lambda_n}u_n(x)u_n^\dagger(y)+S_{Stoch}(x,y)\;.
 \end{equation}
The resulting
full ($k=0$) and reduced ($k>0$) quark propagators are then  used in the construction of the meson
correlation functions.

\subsection{Meson observables}

We use the standard variational approach 
\cite{Michael:1985ne,Luscher:1990ck,Blossier:2009kd}. Our basis of operators is enlarged by
introducing an exponential type of smearing at source/sink.
We construct the cross-correlation matrices  
\begin{equation}
C_{ij}(t)=\langle 0|\mathcal{O}_i(t)\mathcal{O}_j^\dagger(0)|0\rangle
\end{equation}
with the size  up to $10\times10$ with a subsequent solution of the 
generalized eigenvalue problem
\begin{equation}
C(t)\vec{\upsilon}_n(t)=\tilde{\lambda}^{(n)}(t)C(t_0)\vec{\upsilon}_n(t)\;.
\end{equation}
The masses of the eigenstates are obtained by identifying the exponential decay of
the eigenvalues  $\tilde{\lambda}^{(n)}(t)$. Such states are extracted for each low-lying mode
truncation level $k$ in a given quantum channel.

\section{Results}
\subsection{ $J=0$ mesons}
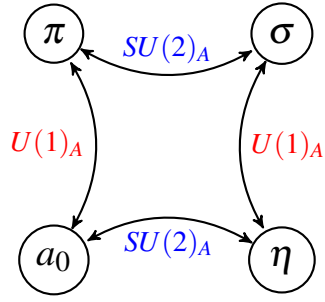
\begin{figure}
\begin{center}
\begin{tikzpicture}[<->,>=stealth',shorten >=1pt,auto,node distance=3cm,
                    thick,main node/.style={circle,draw,font=\sffamily\Large\bfseries}]
  \node[main node] (1)              {$\pi$};
  \node[main node] (2) [right of=1] {$\sigma$};
  \node[main node] (3) [below of=2] {$\eta$};
  \node[main node] (4) [below of=1] {$a_0$};
  \path[every node/.style={font=\sffamily}]
    (1) edge [bend right]  node[blue, pos=0.5, sloped, above] {$SU(2)_A$}(2)
    (2) edge [bend right]  node[red, pos=0.5,  right]         {$U(1)_A$} (3)
    (3) edge [bend right]  node[blue, pos=0.5, sloped, below] {$SU(2)_A$}(4)
    (4) edge [bend right]  node[red, pos=0.5, left]           {$U(1)_A$} (1);
\end{tikzpicture}
 \end{center}
 \caption{Symmetry relations between $J=0$ mesons}\label{fig:sym}
\end{figure}

   \begin{figure}[t]   
\includegraphics[scale=0.58]{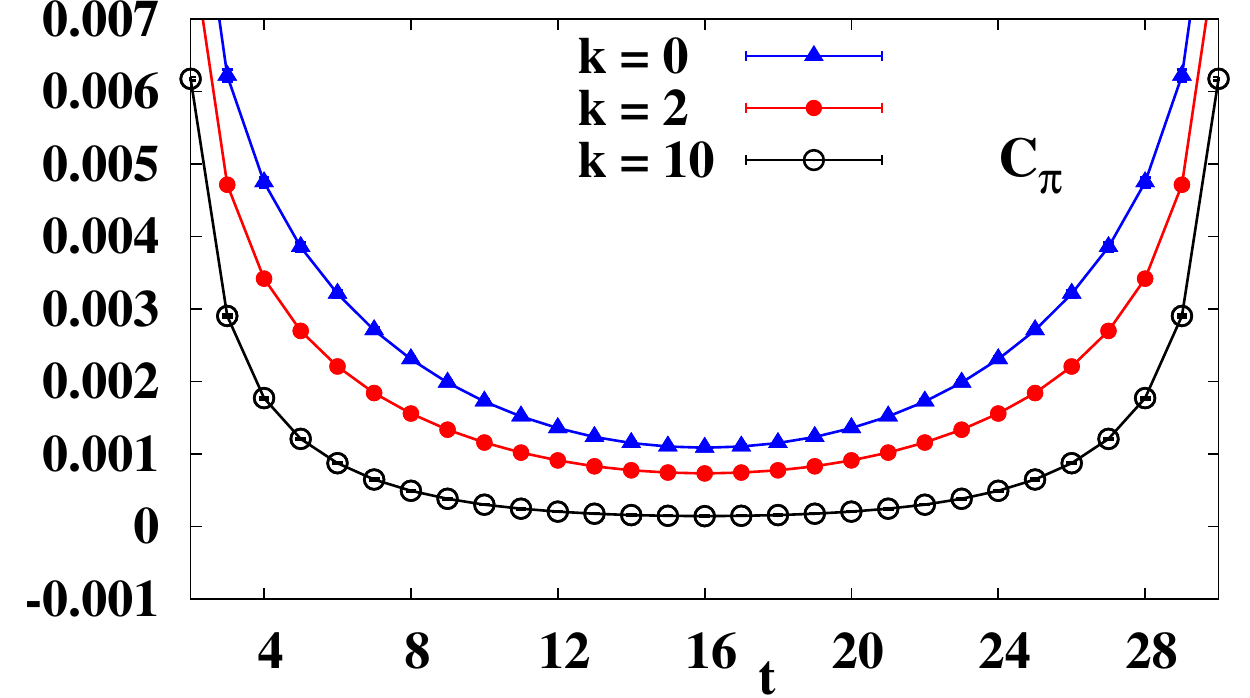}
\includegraphics[scale=0.58]{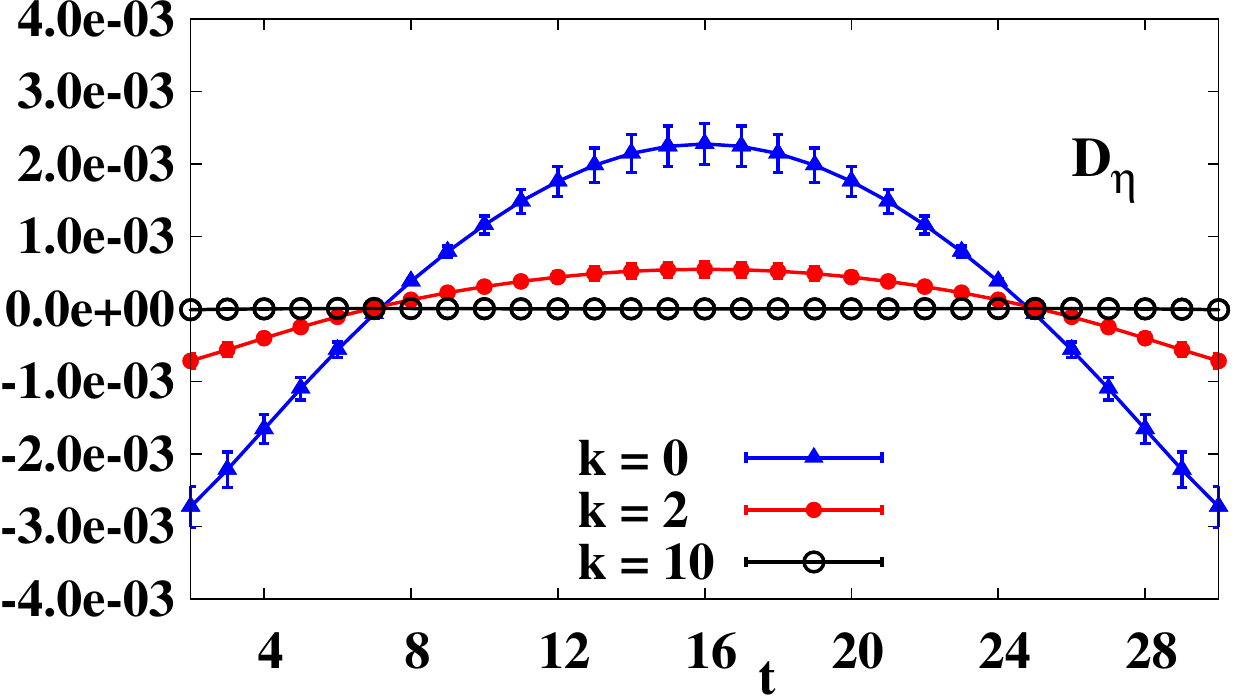}
\caption{ Connected (left) and disconnected (right) parts of the $\eta$
correlator upon the low-mode exclusion.}\label{fig:etacorr}
\end{figure}

The $J=0$ operators and the corresponding representations of the
chiral group are listed in Table \ref{tab:int}. They are connected
by the symmetry transformations that are illustrated in Fig. \ref{fig:sym}
\cite{Glozman:2003bt,COHEN}.

The isovector correlation functions contain only  connected ($C$) 
contributions. Both the connected and disconnected ($D$) contributions
are important for the isoscalar correlators.
In our case we have two degenerate quark flavours, i.e., there is no 
distinction between the $u$ and $d$ quark propagators. Hence, the connected part of the $\sigma$
correlator is identical to the $a_0$ correlator, the same argument applies to $\eta$ and $\pi$:
\begin{equation}
F_{\eta(\sigma)}=C_{\pi(a_0)}+D_{\eta(\sigma)}\;,
\end{equation}
where $F$ represents the full correlator with  given quantum numbers.

In Fig. \ref{fig:etacorr} we show the connected and disconnected contributions in
the $\eta$ channel. In the untruncated
case $k=0$ they are both equally important. After elimination of $k\sim10$ lowest eigenmodes the disconnected contribution
essentially vanishes. 
 
   \begin{figure}[t]
\includegraphics[scale=0.55]{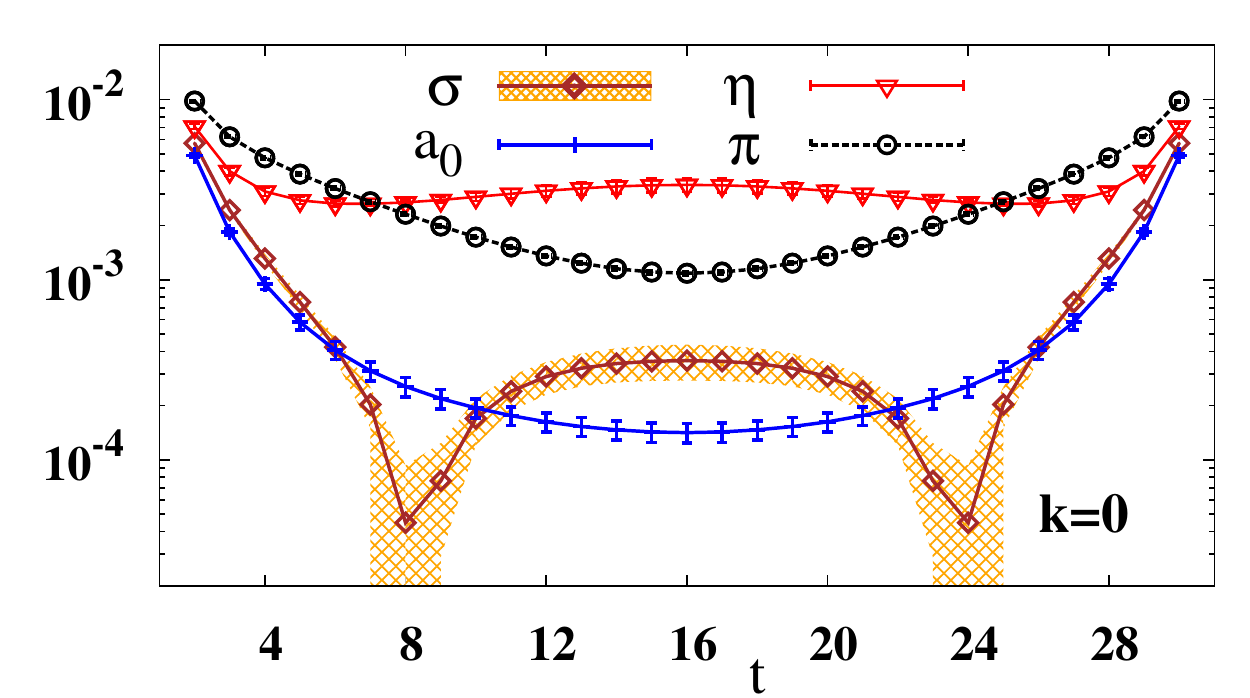}
\includegraphics[scale=0.55]{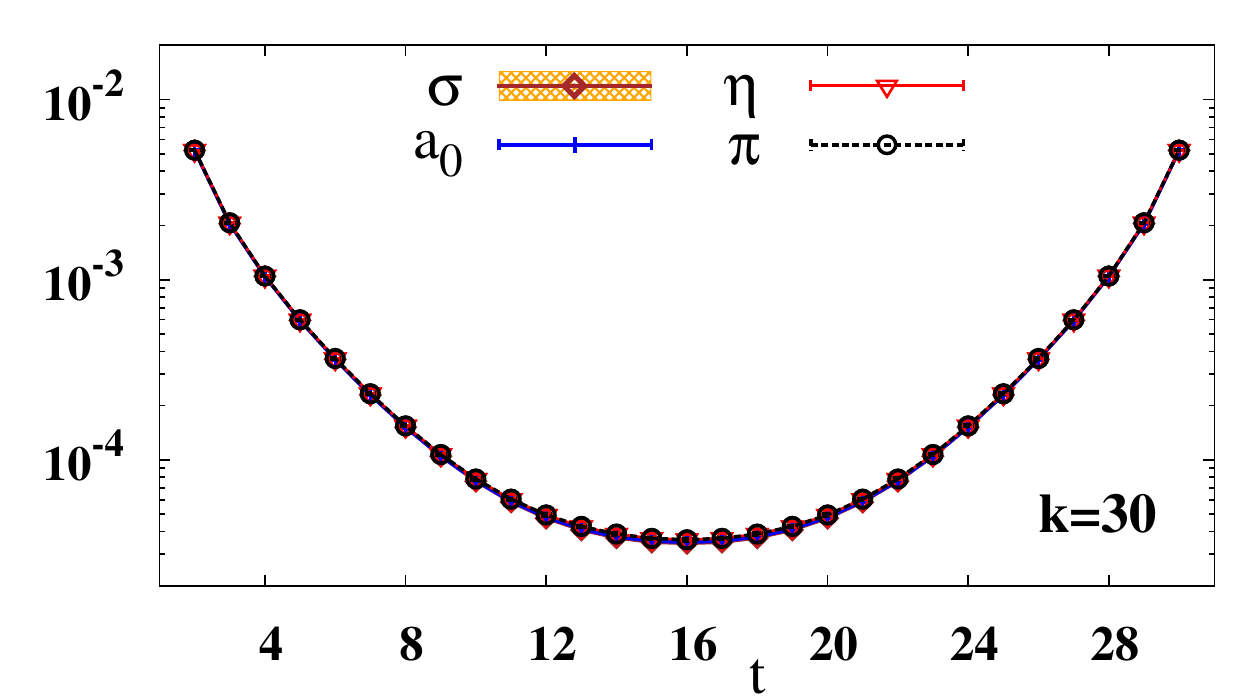}
 \caption{Two-point correlators of $\pi$, $\eta$, $\sigma$, $a_0$ mesons.}\label{fig:mescorr}
\end{figure}
\begin{figure}[t]
\includegraphics[scale=0.46]{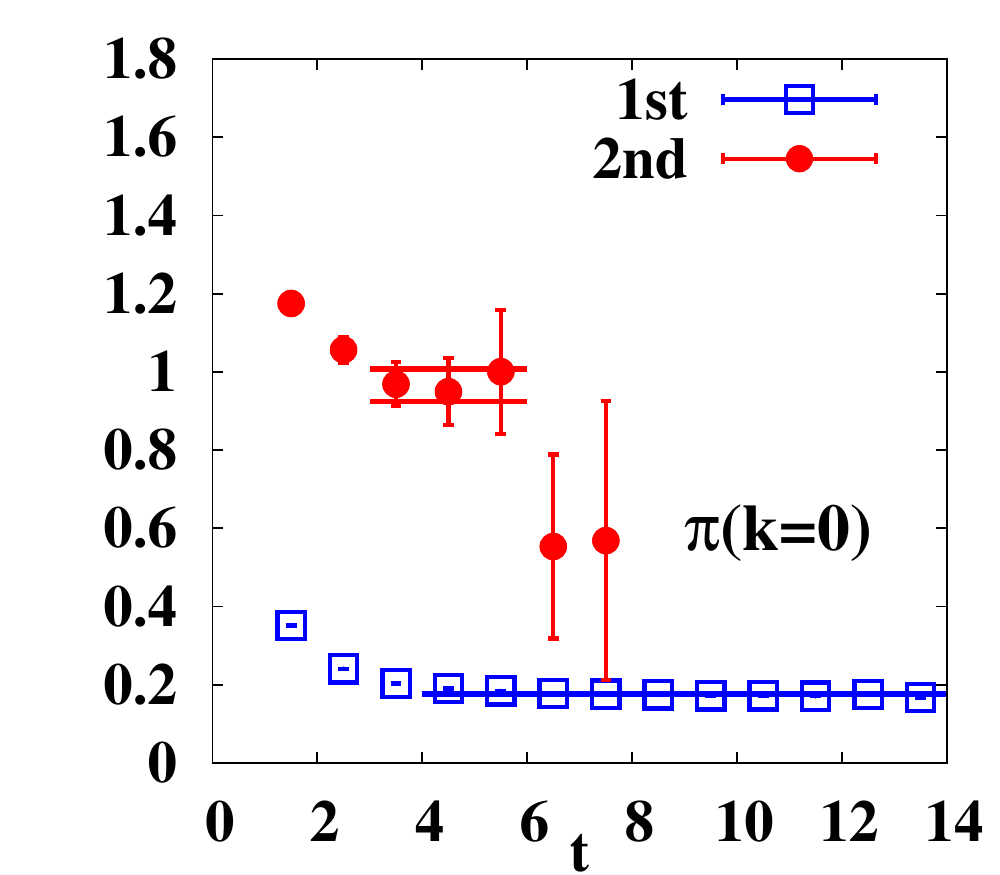}
\includegraphics[scale=0.46]{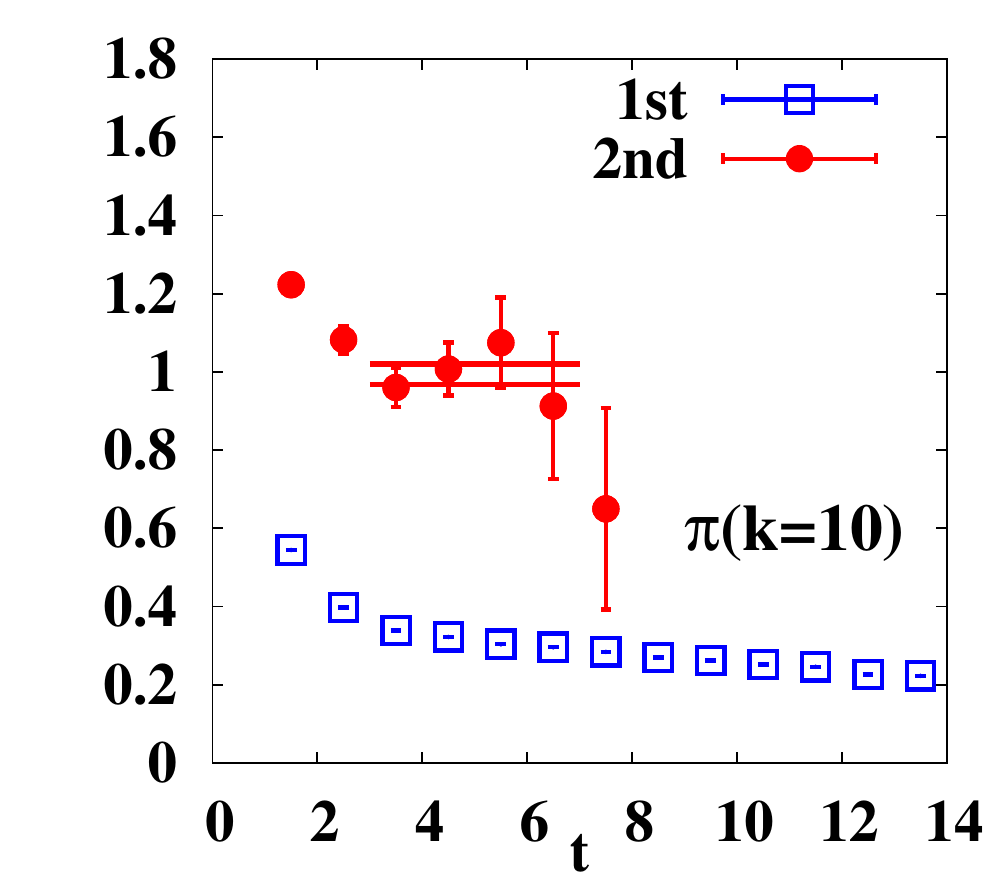}
\includegraphics[scale=0.46]{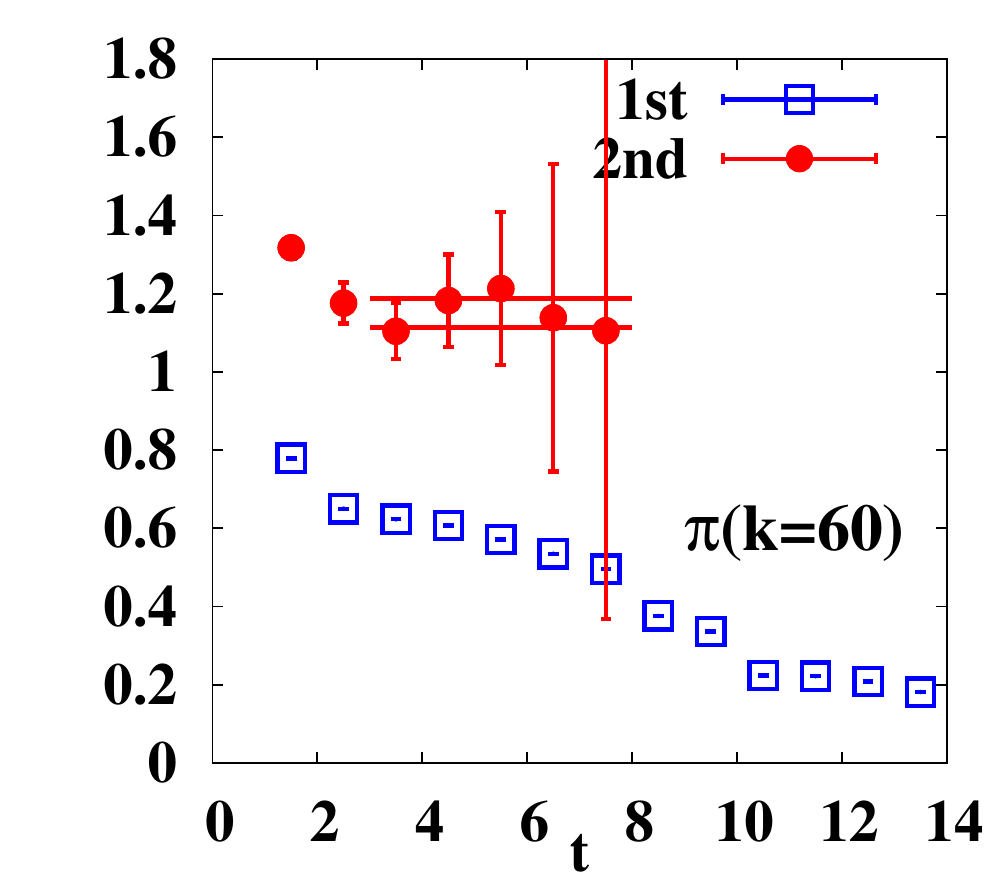}
 \caption{$\pi$ effective masses at $k=0,10,60$. }\label{fig:pimass}
\end{figure}

Figure \ref{fig:mescorr} shows  $\pi$, $\sigma$, $a_0$, $\eta$ point-to-point correlators 
at different truncation levels $k$. 
The $SU(2)_L \times SU(2)_R$ and $U(1)_A$
 symmetries are strongly broken at $k=0$, which makes all four point-to-point
 correlators very different.
 
Upon elimination of a small amount of lowest-lying Dirac eigenmodes
 all correlators become identical. We conclude that both $SU(2)_L \times SU(2)_R $ and $U(1)_A$
symmetries get simultaneously restored. The same quasi-zero modes are responsible for
both symmetry breakings which is consistent with the instanton  induced
mechanism.

 \begin{wraptable}{r}{0.43\textwidth}
\begin{center}
\begin{tabular}{| c | c | c| } 
\hline\hline

   $I,J^{PC}$            & $\mathcal{O}$  & $R$\\ \hline
   $\pi\;(1,0^{-+})$     & $\bar{q}\gamma_5 \frac{\tau}{2}q$& $(1/2,1/2)_a$\\
   $\eta\;(0,0^{-+})$    &  $\bar{q}\gamma_5 q$ &$(1/2,1/2)_b$\\   
   $a_0(1,0^{++})$       &  $\bar{q}\frac{\tau}{2} q$&$(1/2,1/2)_b$  \\
   $\sigma\;(0,0^{++})$  &  $\bar{q} q$&   $(1/2,1/2)_a$            \\
 \hline
\multicolumn{1}{|c|}{\multirow{2}{*}{$\rho(1,1^{--})$}}  
                         &$\bar{q}\gamma_i\frac{\tau}{2}q$           &$(1,0)\oplus (0,1)$\\ 
                         & $\bar{q}\gamma_i\gamma_t\frac{\tau}{2}q$  &$(1/2,1/2)_b$\\ 
 \hline
\multicolumn{1}{|c|}{\multirow{2}{*}{$\omega(0,1^{--})$}}
                         &  $\bar{q}\gamma_iq$                       & $(0,0)$\\ 
                         & $\bar{q}\gamma_i\gamma_tq$                &$(1/2,1/2)_a$\\ 
 \hline
        $a_1(1,1^{++})$  & $\bar{q}\gamma_i \gamma_5\frac{\tau}{2}q$   &$(1,0) \oplus (0,1)$\\ 
        $f_1(0,1^{++})$  & $\bar{q}\gamma_i\gamma_5q $     &$(0,0)$\\ 
        $b_1(1,1^{+-})$  & $\bar{q} \gamma_i \gamma_j \frac{\tau}{2}q$ &$(1/2,1/2)_a$ \\ 
        $h_1(0,1^{+-})$  & $\bar{q} \gamma_i\gamma_jq$    & $(1/2,1/2)_b$              \\  
\hline
 \end{tabular}
 \end{center}
 \caption{$J=0$ and $J=1$ meson interpolating fields $\mathcal{O}$: $R$ denotes an index of the
chiral multiplet within each $J$ \cite{Glozman:2003bt}.}\label{tab:int}
\end{wraptable}

The next question is whether the $J=0$ ground states still
persist after unbreaking. We consider the pseudo-scalar channel 
because  the
original $\pi$ states can be easily identified in the untruncated case.
The effective mass for $\pi$ is shown on Fig. \ref{fig:pimass}. For the
untruncated case $k=0$ we see a very good effective mass plateau and
consequently a clean exponential decay of the correlator. 
Upon exclusion of a small amount of the lowest-lying Dirac modes
the correlation function decays with time no longer exponentially indicating the absence of a
``physical'' state. The unbreaking removes the pion from the physical spectrum,
which is consistent with its Goldstone nature: The quasi-zero modes are
crucially important for its existence.
However, the first excited state of the pion
might survive the truncation. 

We have demonstrated earlier that both $SU(2)_L \times SU(2)_R $ and $U(1)_A$ symmetries 
get simultaneously restored upon unbreaking.
Hence the ground states
of $\sigma, a_0, \eta$ mesons should disappear from the spectrum simultaneously
with the pion. The opposite would contradict the restoration of both symmetries.
 
\begin{figure}[t]
\includegraphics[scale=0.55]{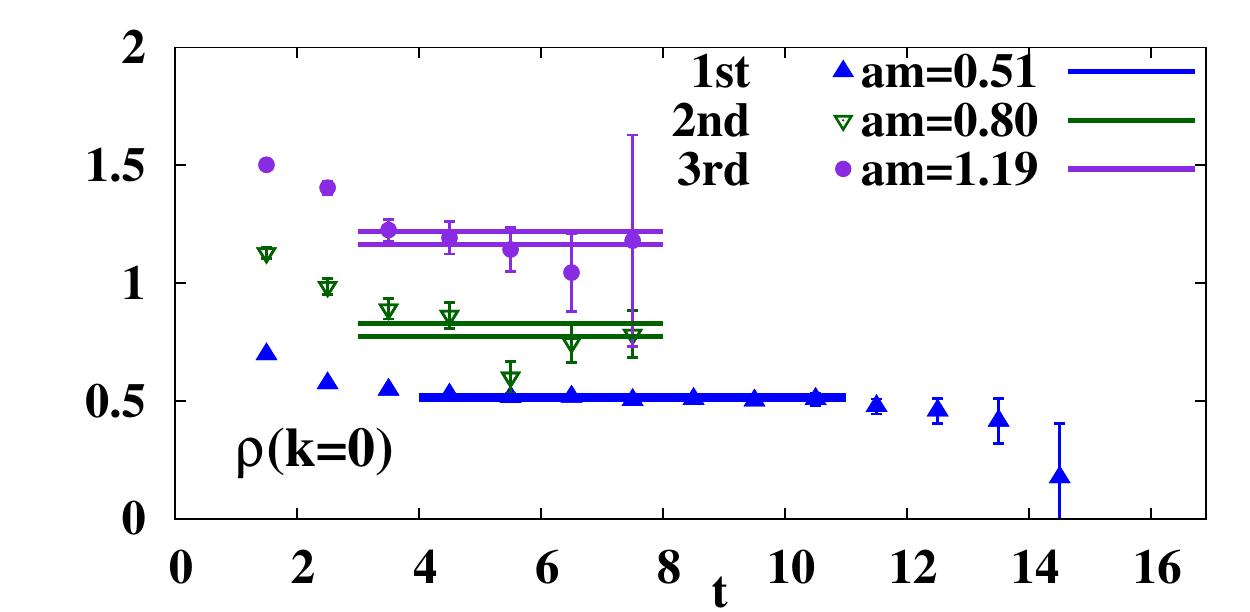} 
\includegraphics[scale=0.55]{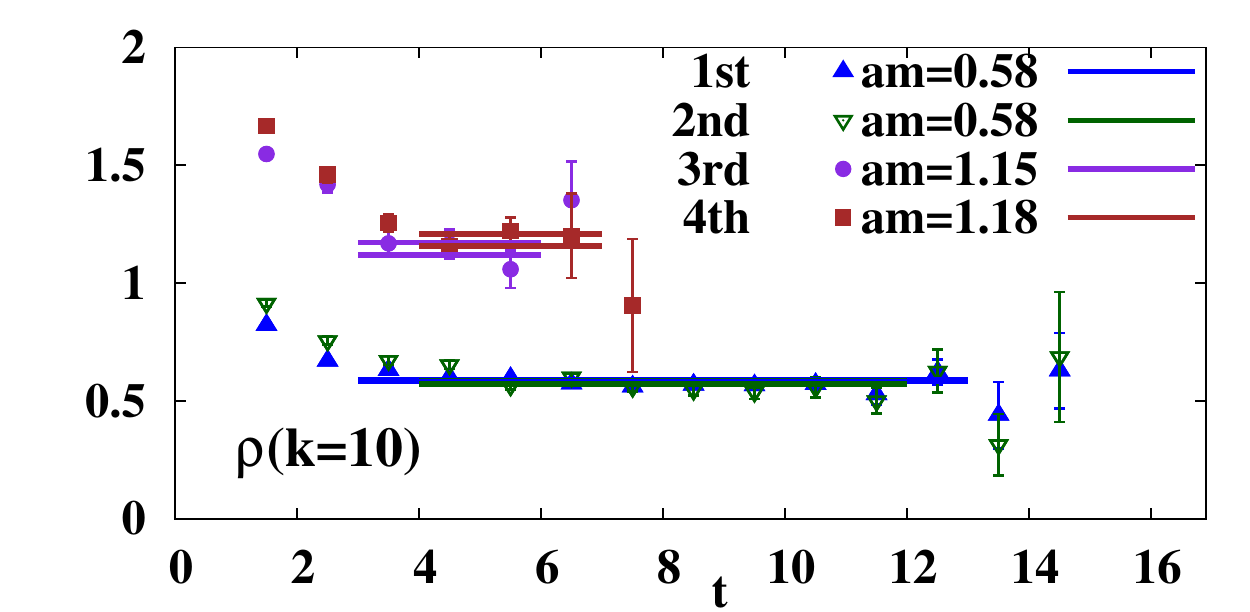}\\
\includegraphics[scale=0.55]{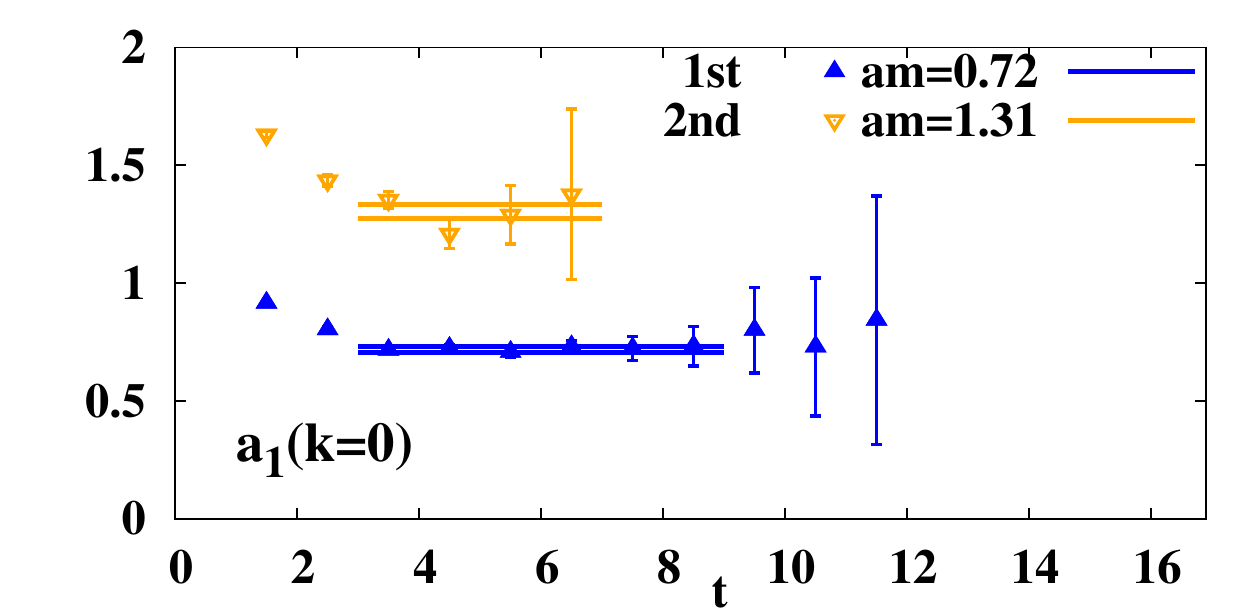} 
\includegraphics[scale=0.55]{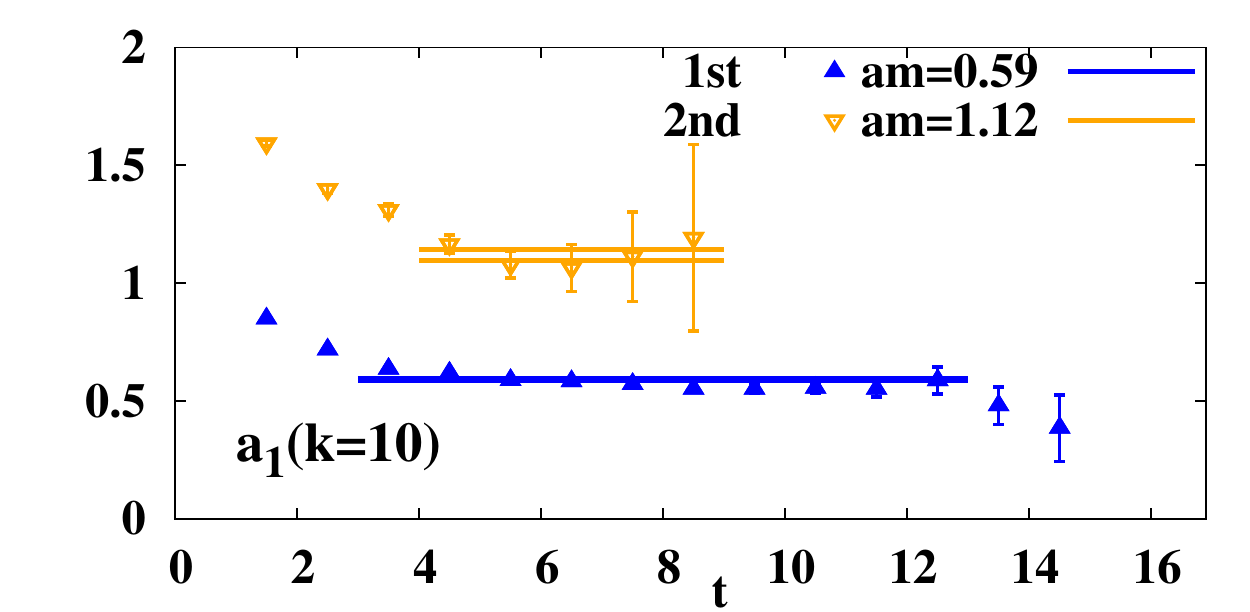}\\
\includegraphics[scale=0.55]{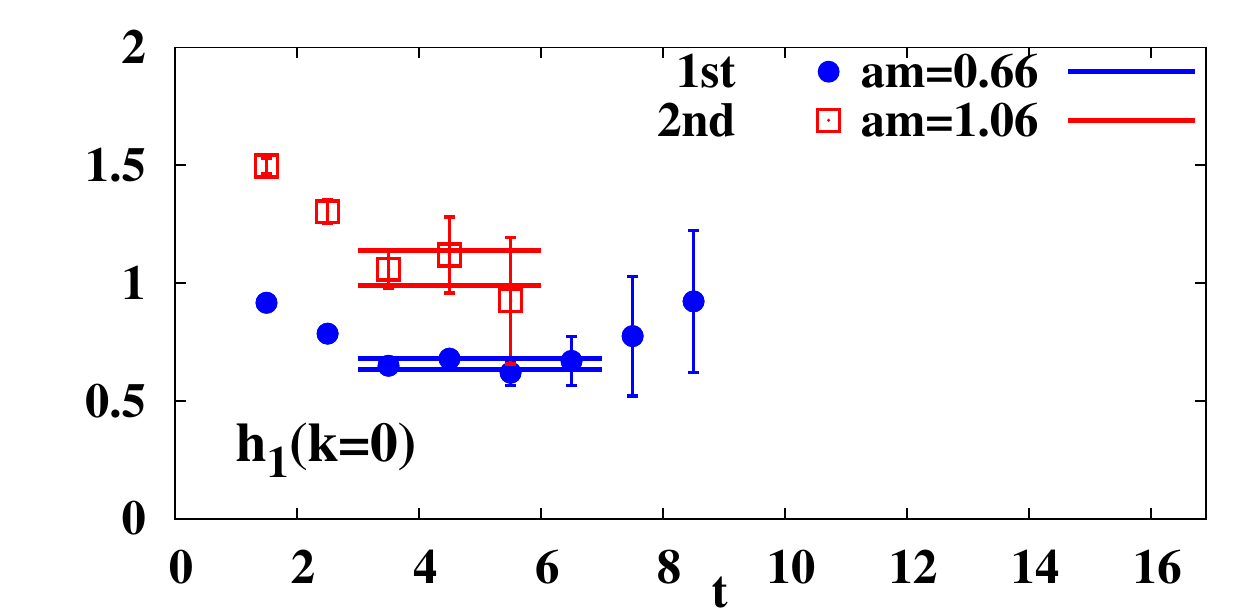} 
\includegraphics[scale=0.55]{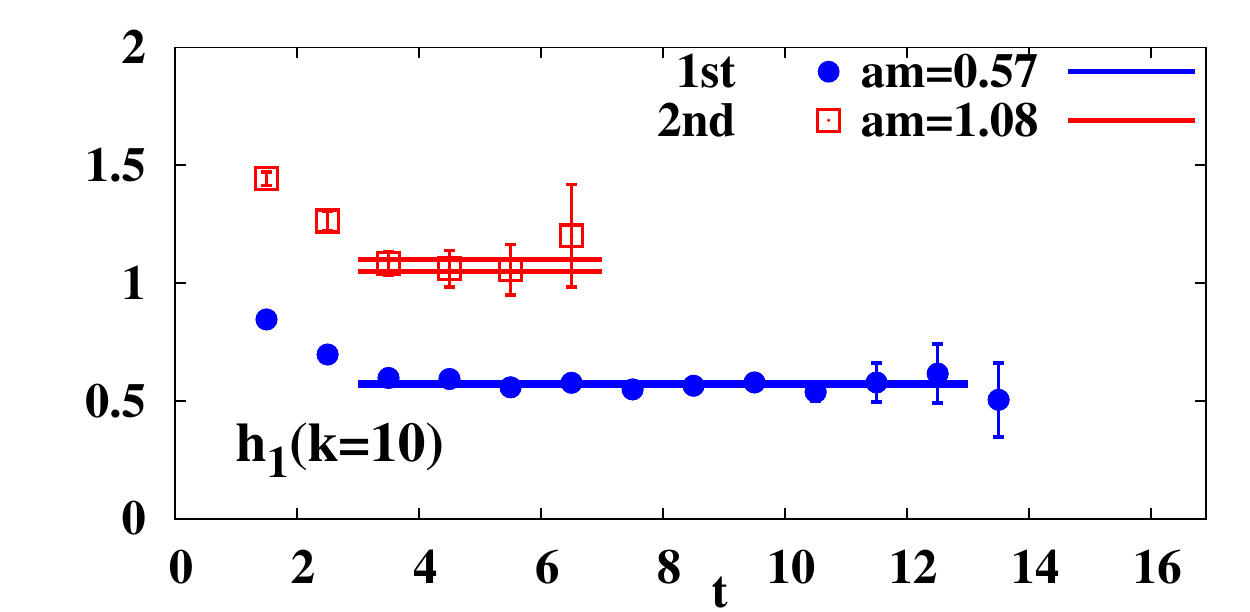}
\caption{$\rho$, $a_1$ and $h_1$ effective masses at k=0,10.} \label{fig:meff}
\end{figure}
\subsection{ $J=1$ mesons}

The interpolating fields and the respective chiral representations
are given in Table \ref{tab:int}. In contrast to the $J=0$ case
the disconnected contributions
to the isoscalar correlators are small compared to the connected ones. The
interpretation
is that the disconnected contributions are supported in the infrared by
the instantons, i.e., by the 't Hooft vertex. This vertex is limited, however,
to the $J=0$ channel. After subtraction of the lowest eigenmodes the small
disconnected contributions in the $J=1$ case vanish.

\begin{figure}[t]
\centering
\includegraphics[scale=0.75]{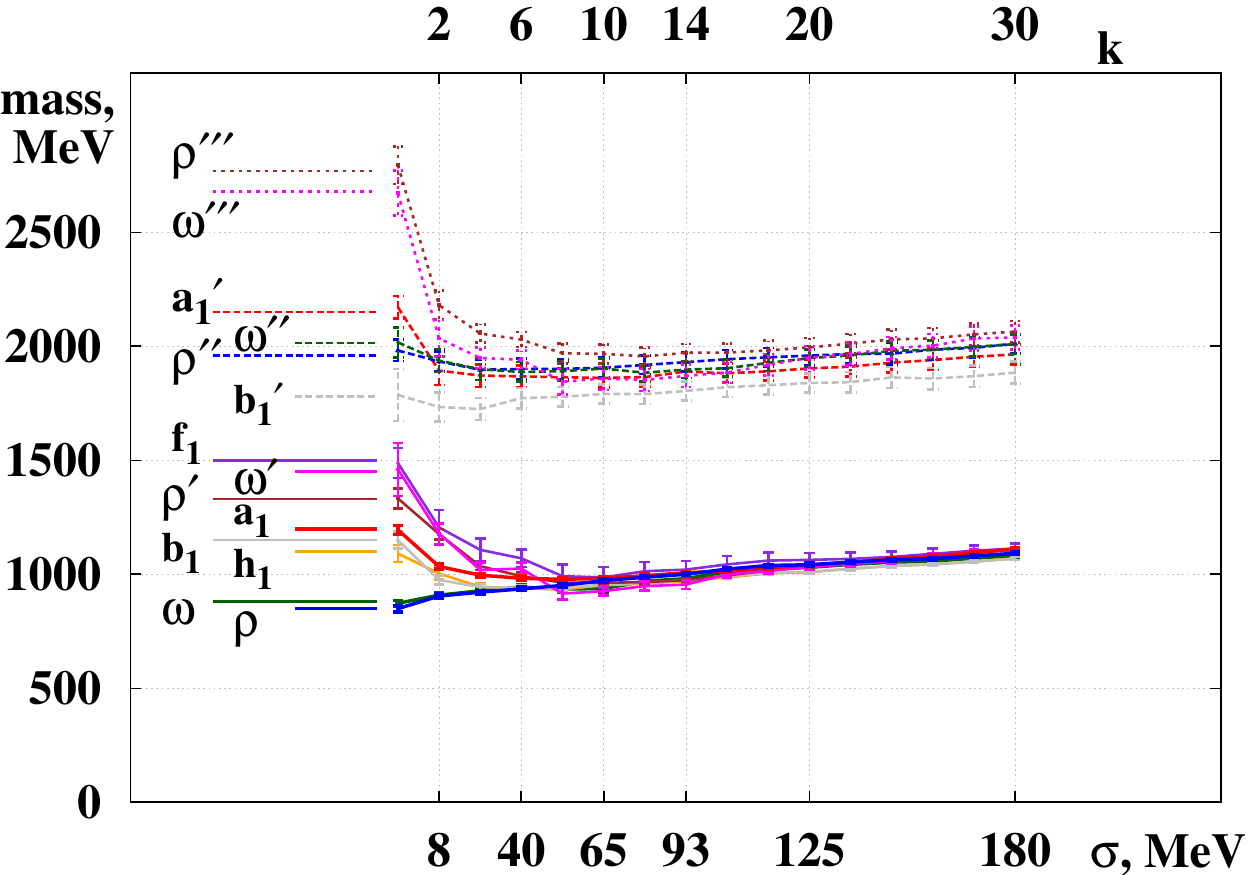}
\caption{Mass evolution of $J=1$ mesons upon exclusion of the low-lying Dirac
modes, $\sigma$ denotes  the energy gap.}\label{fig:allJ1}
 \end{figure}

All $J=1$  states survive the unbreaking, see Fig.
\ref{fig:meff}: After elimination of the lowest-lying Dirac eigenmodes
we observe a clean exponential decay of the corresponding eigenvalues. 
The evolution of the $J=1$ meson 
masses is shown on Fig. \ref{fig:allJ1}. All possible states get degenerate
upon elimination of $\sim 10$ modes. This indicates 
some larger symmetry that includes $SU(2)_L \times SU(2)_R
\times U(1)_A$ as a subgroup. This symmetry can be reconstructed and turns out
to be a $SU(4)$ \cite{Glozman:2014mka}, mixing components of the fundamental
four-component vector $(u_L,u_R,d_L,d_R)$. This symmetry is not a symmetry of
the QCD Lagrangian but should be considered as an emergent symmetry that
appears from the QCD dynamics upon elimination of the quasi-zero Dirac
eigenmodes. It operates only for $J \geq 1$ states. From this symmetry
it is possible to conclude that after truncation
there is no color-magnetic field in the system suggesting that we observe
quantum levels of a dynamical QCD-like string.

\section{Acknowledgements}

We thank S. Aoki, S. Hashimoto and T. Kaneko for supplying us
with the JLQCD gauge configurations and quark propagators and their help and
hospitality.  The calculations were performed on computing clusters of the University of Graz (NAWI Graz).
Support from the
Austrian Science Fund (FWF) through the grants DK W1203-N16 and P26627-N16 
is acknowledged.

\end{document}